\documentclass[twocolumn,english,amsmath,superscriptaddress,prb,aps]{revtex4}
\usepackage[T1]{fontenc}
\usepackage[latin9]{inputenc}
\usepackage{bm}
\usepackage{amsbsy}
\usepackage{graphicx}
\usepackage{esint}

\makeatletter
\@ifundefined{textcolor}{}
{%
 \definecolor{BLACK}{gray}{0}
 \definecolor{WHITE}{gray}{1}
 \definecolor{RED}{rgb}{1,0,0}
 \definecolor{GREEN}{rgb}{0,1,0}
 \definecolor{BLUE}{rgb}{0,0,1}
 \definecolor{CYAN}{cmyk}{1,0,0,0}
 \definecolor{MAGENTA}{cmyk}{0,1,0,0}
 \definecolor{YELLOW}{cmyk}{0,0,1,0}
}

\usepackage{amsthm}

\usepackage{amsfonts}

\@ifundefined{definecolor}{\@ifundefined{definecolor}
 {\usepackage{color}}{}
}{}

\newcommand{\be}{\begin{equation}}
\newcommand{\ee}{\end{equation}}
\newcommand{\bea}{\begin{eqnarray}}
\newcommand{\eea}{\end{eqnarray}}

\renewcommand{\epsilon}{\varepsilon}

\renewcommand{\cite}[1]{[\onlinecite{#1}]}

\usepackage{babel}

\usepackage{babel}

\usepackage{babel}

\makeatother

\usepackage{babel}
\begin{document}

\title{Broken translational symmetry in an emergent paramagnetic phase of
graphene}

\author{Gia-Wei Chern}

\affiliation{Department of Physics, University of Wisconsin-Madison, Madison,
WI 53706, USA}

\affiliation{Theoretical Division, Los Alamos National Laboratory, Los Alamos,
NM, 87545, USA}

\author{Rafael M. Fernandes}

\affiliation{Department of Physics, Columbia University, New York, New York 10027,
USA}

\affiliation{Theoretical Division, Los Alamos National Laboratory, Los Alamos,
NM, 87545, USA}

\author{Rahul Nandkishore}

\affiliation{Department of Physics, Massachusetts Institute of Technology, Cambridge,
MA 02139, USA}

\author{Andrey V. Chubukov}

\affiliation{Department of Physics, University of Wisconsin-Madison, Madison,
WI 53706, USA}
\begin{abstract}
We show that the spin-density wave state on the 
 partially filled honeycomb and triangular lattices is preempted by
a paramagnetic phase that breaks an emergent $Z_{4}$ symmetry of
the system, associated with the four inequivalent arrangements of
spins in the quadrupled unit cell. Unlike other emergent paramagnetic
phases in itinerant and localized-spin systems, this state preserves
the rotational symmetry of the lattice but breaks its translational
symmetry, giving rise to a super-lattice structure that can be detected
by scanning tunneling microscopy. This emergent phase also has distinctive
signatures in the magnetic spectrum that can be probed experimentally.
\end{abstract}
\maketitle

\section{Introduction}

Unconventional paramagnetic phases are characterized not only by the
absence of long-range spin order, but also by a broken symmetry related
to new degrees of freedom that emerge from the collective magnetic
behavior of the system. As a result, their elementary excitations
and thermodynamic properties are rather different than those of an
ordinary paramagnet. These phases usually appear in frustrated systems
with localized spins, as a result of the interplay between frustration
and fluctuations. Canonical examples include the Ising-nematic phase
of the extended Heisenberg model on the square lattice \cite{Chandra},
the spin-nematic phase of the Heisenberg model on the kagome lattice
\cite{Chalker}, and the magnetic-charge ordered phase in kagome spin
ice \cite{kagome-ice}. Itinerant magnetic systems can also display
paramagnetic phases with unusual broken symmetries. This is believed
to be the case in the ruthenates~\cite{ruthenates} and in the iron-based
superconductors~\cite{Chu10,fernandes,Eremin}. In these systems
the emergent paramagnetic phase breaks the lattice rotational symmetry,
while the spin-rotational and lattice translational symmetries remain
preserved.

\begin{figure}
\includegraphics[width=0.75\columnwidth]{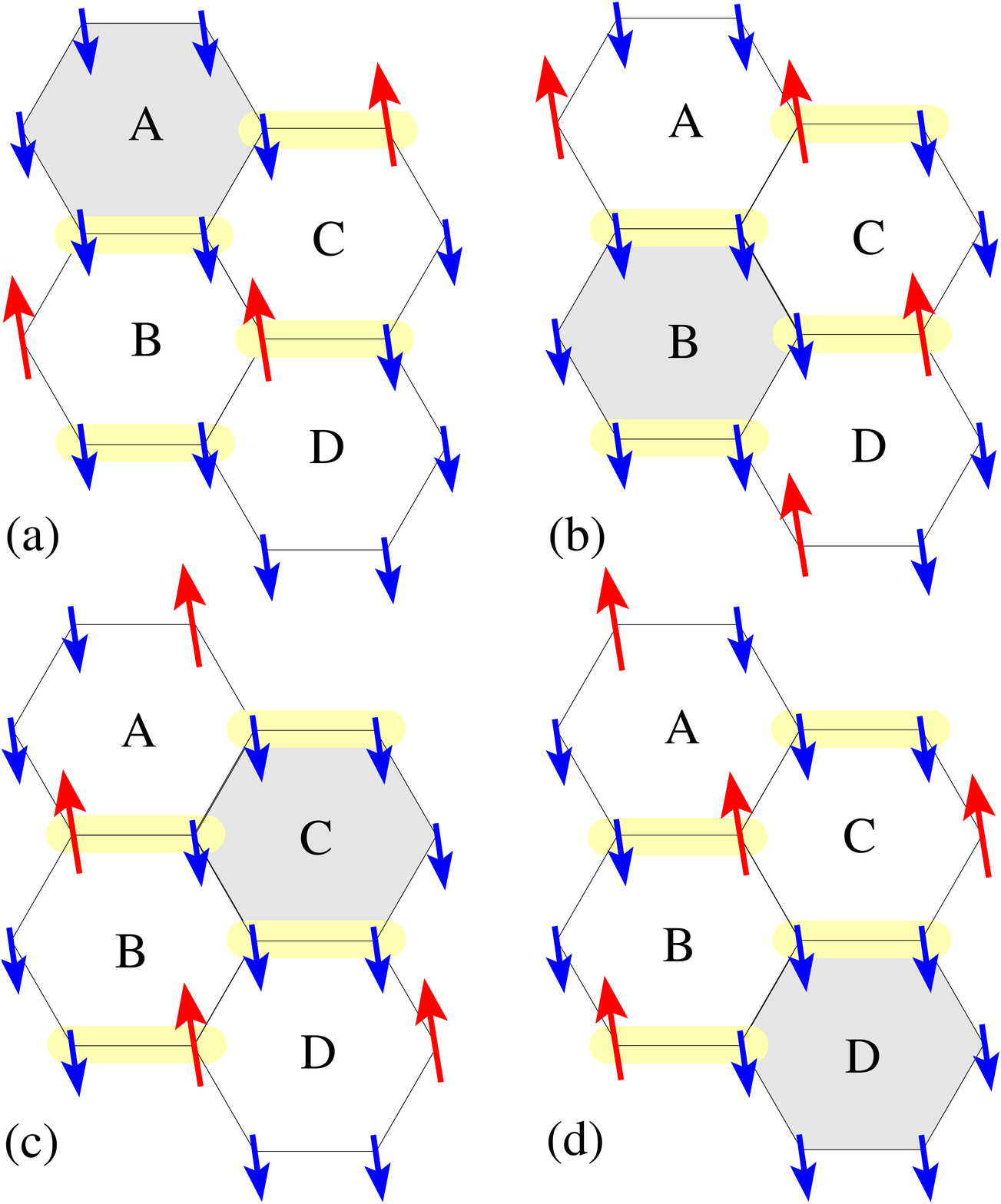} \caption{(Color online) The four inequivalent uniaxial SDW states. The quadrupled
unit cell is highlighted with yellow boxes. Among the eight sites
of the enlarged unit cell, two have large spin moment $\left\langle \mathbf{S}_{i}\right\rangle =3\boldsymbol{\Delta}$
(red arrows) and six have small moment $\left\langle \mathbf{S}_{i}\right\rangle =-\boldsymbol{\Delta}$
(blue arrows). The total spin in each unit cell is zero. The four
different states correspond to the additional $Z_{4}$ symmetry of
the order parameter manifold. \label{fig:sdw3q}}
\end{figure}

In this paper, we present an unusual itinerant paramagnetic phase
that breaks the translational invariance without changing the point-group
symmetry of the lattice. This phase arises in partially filled hexagonal
(triangular and honeycomb) lattices, preempting a spin-density wave
(SDW) order, and could potentially be realized in single-layer graphene
doped near the saddle point of the band-structure ($3/8$ or $5/8$
filling)~\cite{CastroNeto,McChesney}. The SDW order below $T_{N}$
for fermions on a hexagonal lattice is uniaxial, with all spins pointing
along the same direction~\cite{Nandkishore12}. The magnetic unit
cell contains eight sites, six of which have moment $-\boldsymbol{\Delta}$
and two have moment $3\boldsymbol{\Delta}$, see Fig.~\ref{fig:sdw3q}.
This state breaks not only the $O(3)$ spin-rotational symmetry, but
also a discrete $Z_{4}$ symmetry related to the four inequivalent
choices for the positions of the large $3\boldsymbol{\Delta}$ spin
moments in the eight-site unit cell. These four inequivalent spin
configurations transform into each other upon translation of the origin
of coordinates to neighboring hexagons - from point $A$ to points
$B,C$ and $D$ in Fig.~\ref{fig:sdw3q}. Thus, breaking the $Z_{4}$
symmetry corresponds to breaking the translational symmetry of the
lattice.

\begin{figure}
\includegraphics[width=0.75\columnwidth]{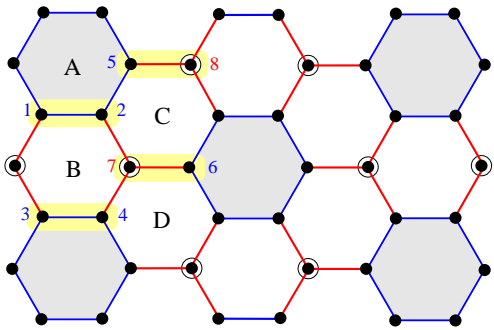} \caption{(Color online) Schematic representation of the emergent $Z_{4}$ symmetry-breaking
phase. The quadrupled unit cell is highlighted with yellow boxes;
the numbers indicate the eight inequivalent sites. This state has
the same broken translational symmetry as that of the SDW in Fig.~1(a),
but $O(3)$ symmetry is preserved (no arrows). The nearest-neighbor
correlation function $\left\langle \mathbf{S}_{i}\cdot\mathbf{S}_{j}\right\rangle $
is $\Delta^{2}$ for blue bonds and $-3\Delta^{2}$ for red bonds.
Other three states are obtained by moving the origin of coordinates
from A to either B or C or D. \label{fig:z4}}
\end{figure}

Of course, once the $O(3)$ symmetry is broken, the $Z_{4}$ symmetry
has to be broken too. We show, however, that the $Z_{4}$ symmetry
breaks down at higher temperatures than the $O(3)$ symmetry. As a
result, the SDW ordering at $T_{N}$ is preempted by a phase transition
at $T_{Z_{4}}>T_{N}$, which falls into the universality class of
the four-state Potts model. In the $Z_{4}$ phase at $T_{N}<T<T_{Z_{4}}$,
$\langle\mathbf{S}_{i}\rangle=0$ for all sites (i.e., this phase
is a paramagnet), and the unit cell is a hexagon (green dashed line
in Fig.~\ref{fig:FS}),
 i.e., the 
 $C_{6}$ rotational symmetry of the lattice is preserved. Yet, the
unit cell has eight inequivalent sites -- for six of them ($i=1...6$
in Fig.~\ref{fig:z4}) the bond correlators $\left\langle \mathbf{S}_{i}\cdot\mathbf{S}_{i+\bm{\delta}}\right\rangle $
with their nearest neighbors are $\Delta^{2}$ and $-3\Delta^{2}$
(blue and red bonds in Fig.~\ref{fig:sdw3q}(b)), while for the remaining
two sites ($i=7,8$) all bond correlators are $-3\Delta^{2}$. The
broken $Z_{4}$ symmetry corresponds to choosing these two ``special''
sites out of the eight sites in the unit cell. One such choice is
shown in Fig. \ref{fig:z4}. One can easily verify that the other
three choices correspond to moving the origin of the coordinates from
$A$ to one of the points $B,C$, or $D$ in Fig. \ref{fig:z4}. This
obviously implies that the the translational symmetry of the lattice
is broken. Experimentally, the quadrupled unit cell in the $Z_{4}$
phase can be readily probed by scanning tunneling microscopy (STM).
Furthermore, we show that the transition to this phase is accompanied
by a jump of the staggered spin susceptibility, which can be probed
by neutron scattering or nuclear magnetic resonance (NMR).

\section{The uniaxial SDW order}

\begin{figure}
\includegraphics[width=0.9\columnwidth]{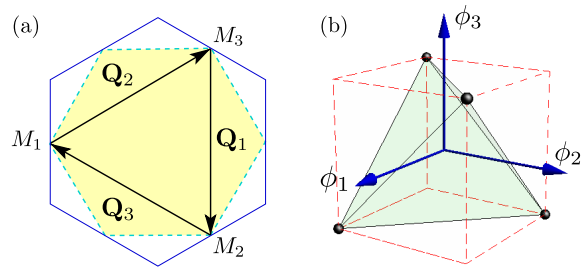}

\caption{(Color online) (a) The Fermi surface at the doping level of interest
is a hexagon inscribed within the hexagonal Brillouin zone (BZ), for
both honeycomb and triangular lattices. The FS has three saddle points
$M_{a}$ located at the corners of the hexagon. Pairs of inequivalent
saddle points are connected by three inequivalent nesting vectors
$\mathbf{Q}_{1}=(0,2\pi/\sqrt{3})$, and $\mathbf{Q}_{2,3}=(\mp\pi/3,-\pi/\sqrt{3})$.
(b) Order-parameter space of the preemptive $Z_{4}$ phase. In the
ordered phase the vector order parameter $\bm{\phi}=(\phi_{1},\phi_{2},\phi_{3})$
points toward one of the four corners of a regular tetrahedron. \label{fig:FS}}
\end{figure}

The Fermi surface (FS) of graphene near $3/8$ or $5/8$ filling is
near-nested and contains three saddle points with nearly vanishing
Fermi velocity (the three $M_{a}$ points in Fig.~\ref{fig:FS}(a)).
Pairs of inequivalent $M_{a}$ points are connected by three commensurate
nesting vectors $\mathbf{Q}_{1}=(0,2\pi/\sqrt{3})$ and $\mathbf{Q}_{2,3}=(\mp\pi/3,-\pi/\sqrt{3})$.
The divergent density of states at the $M$-points makes doped graphene
a fertile ground for exploring nontrivial many-body density-wave and
superconducting states \cite{Nandkishore,TaoLi,MoraisSmith,FaWang,Vozmediano,Thomale,Gonzalez}.
The SDW instability is subleading to a chiral d-wave superconductivity
exactly at $3/8$ or $5/8$ filling~\cite{Nandkishore}, but can
become the leading instability slightly away from $3/8$ or $5/8$
filling~\cite{Thomale,Nandkishore12}.

 In particular, the FS at the saddle-point doping levels, e.g. $3/8$
or $5/8$ for the honeycomb lattice, is a perfect hexagon inscribed
within a hexagonal Brillouin zone (BZ) as shown in Fig.~\ref{fig:FS}(a).
This FS is completely nested by three wavevectors $\mathbf{Q}_{1}=(0,2\pi/\sqrt{3})$,
and $\mathbf{Q}_{2,3}=(\pm\pi/3,-\pi/\sqrt{3})$, and the nesting
opens the door to an SDW instability. However, not all points on the
Fermi surface are of equal importance. In particular, the three saddle
points $M_{a}$ ($a=1,2,3$) give rise to a logarithmic singularity
in the DOS and control the SDW instability at weak coupling.

Thus, we consider the following Hamiltonian:
\begin{eqnarray}
 &  & H=\sum_{a=1,2,3}\epsilon_{a}c_{a,\alpha}^{\dagger}c_{a,\alpha}^{\phantom{dag}}\\
 &  & \quad-\sum_{a\neq b}\left(g_{2}\, c_{a,\alpha}^{\dagger}c_{b,\beta}^{\dagger}c_{b,\beta}^{\phantom{\dag}}c_{a,\alpha}^{\phantom{\dag}}+g_{3}\, c_{a,\alpha}^{\dagger}c_{a,\beta}^{\dagger}c_{b,\beta}^{\phantom{\dag}}c_{b,\alpha}^{\phantom{\dag}}\right), \nonumber
 \label{eq_SDW_Hamilt}
\end{eqnarray}
 where $c_{a,\alpha}^{\dagger}$ creates electrons with spin $\alpha$
around the saddle point $M_{a}$. There are two electron-electron
interactions that contribute to the SDW channel, 
namely $g_{2}$ and $g_{3}$, which represent the forward and umklapp
scatterings, respectively. The dispersions in the vicinity of the
saddle points are
\begin{eqnarray}
 \epsilon_{1}(\mathbf{k})&=&\frac{3t_{1}}{4}(k_{y}^{2}-3k_{x}^{2}), \\
\epsilon_{2,3}(\mathbf{k})&=& -\frac{3t_{1}}{4}2k_{y}(k_{y}\mp\sqrt{3}k_{x}),  
\label{eq:dispersion}
\end{eqnarray}
 where $t_{1}$ is the nearest-neighbor hopping constant. The quartic
interaction terms in Eq.~(\ref{eq_SDW_Hamilt}) can be decoupled
via the Hubbard Stratonovich transformation with the SDW order parameters:
$\bm{\Delta}_{i}=\bm{\Delta}_{a,b}=\frac{g_{2}+g_{3}}{3}\sum_{\mathbf{k}}\langle c_{a,\alpha}^{\dagger}\bm{\sigma}_{\alpha\beta}^{\phantom{\dagger}}c_{b,\,\beta}^{\phantom{\dagger}}\rangle$.
Each of these vector order parameters corresponds to a nesting vector
which connects two saddle points: $\mathbf{Q}_{i}=\mathbf{M}_{a}-\mathbf{M}_{b}$.
The partition function of the system can then be written as $Z=\int\mathcal{D}c^{\dagger}\mathcal{D}c^{\phantom{\dag}}\mathcal{D}\mathbf{\Delta}_{i}\,\exp(-S[c^{\dagger},c^{\phantom{\dag}},\bm{\Delta}])$
with
\begin{eqnarray}
 &  & S[c,c^{\dagger},\bm{\Delta}_{i}]=\sum_{a}\int_{\tau}c_{a,\alpha}^{\dagger}(\partial_{\tau}-\epsilon_{a})c_{a,\alpha}\\
 &  & \quad+\frac{2}{g_{2}+g_{3}}\sum_{i}\int_{x,\tau}|\bm{\Delta}_{i}|^{2}-\sum_{a\neq b}\int_{\tau}\bm{\Delta}_{a,b}\cdot c_{a,\alpha}^{\dagger}\bm{\sigma}_{\alpha\beta}^{\phantom{\dagger}}c_{b,\,\beta}^{\phantom{\dagger}},\nonumber
\end{eqnarray}
 where $\int_{\tau}=\int_{0}^{1/T}d\tau$. The fermionic part becomes
quadratic and can be integrated out. By expanding the resulting action
to fourth order in $\bm{\Delta}_{i}$, we obtain the effective action:
\begin{eqnarray}
 &  & \!\!\!\!\!\!\! S[\bm{\Delta}_{i}]=r_{0}\sum_{i}\int_{x}|\bm{\Delta}_{i}|^{2}+\frac{u}{2}\int_{x}\left(|\bm{\Delta}_{1}|^{2}\!+\!|\bm{\Delta}_{2}|^{2}\!+\!|\bm{\Delta}_{3}|^{2}\right)^{2}\nonumber \\
 &  & \!\!\!\!+\frac{v}{2}\int_{x}\Bigl[\left(|\bm{\Delta}_{1}|^{2}\!+\!|\bm{\Delta}_{2}|^{2}\!-\!2|\bm{\Delta}_{3}|^{2}\right)^{2}\!+\!3\left(|\bm{\Delta}_{1}|^{2}\!-\!|\bm{\Delta}_{2}|^{2}\right)^{2}\Bigr]\nonumber \\
 &  & \!\!\!\!-\frac{g}{2}\int_{x}\Bigl[(\bm{\Delta}_{1}\cdot\bm{\Delta}_{2})^{2}+(\bm{\Delta}_{2}\cdot\bm{\Delta}_{3})^{2}+(\bm{\Delta}_{3}\cdot\bm{\Delta}_{1})^{2}\Bigr]\!+\cdots\,\,\,\quad\label{S_eff}
\end{eqnarray}

Here $r_{0}\propto(T-T_{N})$, where $T_{N}$ is the mean-field SDW
transition temperature. The coefficients $u,v,g$ in Eq.~\ref{S_eff}
were calculated in Ref.~\cite{Nandkishore12} and found to be positive,
with $v/u=1/\log\left(W/T_{N}\right)\ll1$ and $g/u=\left(T_{N}/W\right)/\log\left(W/T_{N}\right)\ll~1$,
where $W$ is the bandwidth.

Minimizing $S[\bm{\Delta}_{i}]$ with respect to $\bm{\Delta}_{i}$
and neglecting momentarily the fluctuations of the $\bm{\Delta}_{i}$
fields, we see that $v>0$ implies that the magnitudes of $\bm{\Delta}_{i}$
are equal, while $g>0$ makes all $\bm{\Delta}_{i}$ collinear. The
particular uniaxial state with $(\bm{\Delta}_{1},\bm{\Delta}_{2},\bm{\Delta}_{3})=(\Delta,\Delta,\Delta)\,\hat{\mathbf{n}}$
is shown in Fig.~\ref{fig:sdw3q}(a). There exists, however, three
other states with the same energy, $(\Delta,-\Delta,-\Delta)\hat{\mathbf{n}}$,
$(-\Delta,\Delta,-\Delta)\hat{\mathbf{n}}$, and $(-\Delta,-\Delta,\Delta)\hat{\mathbf{n}}$.
These states can \emph{not} be obtained from the one shown in Fig.~\ref{fig:sdw3q}(a)
by a global spin rotation. Instead, these four degenerate states are
related by a translational $Z_{4}$ symmetry -- they transform into
each other by moving the origin of coordinates from $A$ to $B,C$,
or $D$ (Fig.~\ref{fig:sdw3q}(b)--(d)). The ground state in Fig.~\ref{fig:sdw3q}(a)
chooses a particular direction of $\hat{\mathbf{n}}$ and also one
of the four positions of the origin of coordinates and therefore breaks
$O(3)\times Z_{4}$ symmetry.

\section{Preemptive $\mathbf{Z_{4}}$ phase}

\subsection{Order parameters}

We now allow ${\bf \Delta}_{i}$ to fluctuate and analyze the possible
emergence of a phase in which $Z_{4}$ symmetry is broken but $O(3)$
symmetry is preserved. In such a phase $\langle\bm{\Delta}_{i}\rangle=0$,
but $\langle\bm{\Delta}_{i}\cdot\bm{\Delta}_{j}\rangle\neq0$. A proper
order parameter for the $Z_{4}$ phase is the triplet $\bm{\phi}=(\phi_{1},\phi_{2},\phi_{3})$,
where $\phi_{i}=g\left(\bm{\Delta}_{j}\cdot\bm{\Delta}_{k}\right)$
and $(ijk)$ are cyclic permutations of $(123)$. The $Z_{4}$ symmetry
breaking phase has $\langle\phi_{i}\rangle=\pm\phi$, with the constraint
$\phi_{1}\phi_{2}\phi_{3}>0$. To investigate whether this state emerges
we go beyond the mean-field approximation for $S[\bm{\Delta}_{i}]$
by including fluctuations of the $\bm{\Delta}_{i}$ fields, and re-express
the action in terms of the collective variables $\phi_{i}$. We analyze
this action assuming that fluctuations of $\phi_{i}$ are weak and
check whether a non-zero $\langle\phi_{i}\rangle$ emerges above the
SDW transition temperature.

To obtain the action in terms of $\phi_{i}$ we apply a Hubbard-Stratonovich
transformation~\cite{fernandes} and introduce six auxiliary fields,
one for each quartic term. These six fields include two fields $\zeta_{1}\propto(\bm{\Delta}_{1}^{2}+\bm{\Delta}_{2}^{2}-2\bm{\Delta}_{2}^{2})$
and $\zeta_{2}\propto(\bm{\Delta}_{1}^{2}-\bm{\Delta}_{2}^{2})$ which
break the $C_{6}$ rotational symmetry, the three fields $\phi_{i}\propto\bm{\Delta}_{j}\cdot\bm{\Delta}_{k}$
associated with the $Z_{4}$ symmetry breaking, and the field $\psi\propto(\bm{\Delta}_{1}^{2}+\bm{\Delta}_{2}^{2}+\bm{\Delta}_{3}^{2})$
associated with the Gaussian fluctuations of the $\bm{\Delta}_{i}$
fields. Details of the Hubbard-Stratonovich transformation can be
found in Appendix~A. In particular, we show that the non-zero values
of $\zeta_{1}$ and $\zeta_{2}$ are energetically unfavorable because
$v>0$ so we set $\zeta_{1}=\zeta_{2}=0$ in the following analysis
and consider states that preserves the lattice rotational symmetry.

The quartic terms in Eq.~(\ref{S_eff}) can be decoupled using the
auxiliary fields $\phi_{i}$ and $\psi$. Because we allow the $\bm{\Delta}_{i}$
fields to fluctuate, we include non-uniform space/time configurations,
i.e., replace $\bm{\Delta}\rightarrow\bm{\Delta}_{q,\omega}$ and
$r_{0}\rightarrow r_{0}+q^{2}+\Gamma|\omega_{m}|$ in Eq.~(\ref{S_eff}),
with $\omega_{m}=2m\pi T$. Near a finite temperature phase transition
thermal fluctuations are the most relevant, and we restrict our analysis
to the $\omega_{m}=0$ component. The new action now depends only
on the $\psi$ and $\bm{\phi}$ \textit{\emph{fields:}}
\begin{equation}
S[\psi,\bm{\phi}]=\int_{x}\left(\frac{|\bm{\phi}|^{2}}{2g}-\frac{\psi^{2}}{2u}\right)+\frac{3}{2}\int_{q}\log\left(\det\mathcal{\hat{X}}\right).\label{S}
\end{equation}
 where $|\bm{\phi}|^{2}=\sum_{i}\phi_{i}^{2}$, $\int_{q}=\frac{VT}{\left(2\pi\right)^{d}}\int d^{d}q$,
and $V$ is the volume of the system. The matrix $\mathcal{\hat{X}}$
is
\begin{eqnarray}
\mathcal{\hat{X}}=\left(\begin{array}{lcr}
\tilde{\chi}_{q}^{-1} & -\phi_{3} & -\phi_{2}\\
-\phi_{3} & \tilde{\chi}_{q}^{-1} & -\phi_{1}\\
-\phi_{2} & -\phi_{1} & \tilde{\chi}_{q}^{-1}
\end{array}\right),\label{K-matrix}
\end{eqnarray}
 with renormalized $\tilde{\chi}_{q}^{-1}=r_{0}+\psi+q^{2}\equiv r+q^{2}$.
In the absence of broken $Z_{4}$ symmetry, long-range SDW order sets
in at $r=0$, hence an intermediate phase exists if $Z_{4}$ symmetry
is broken at some $r>0$.

The action (\ref{K-matrix}) is an unconstrained function of $\psi$,
which is the usual situation for Gaussian fluctuations~\cite{Chaikin},
and reflects the fact that $\langle\boldsymbol{\Delta}_{i}^{2}\rangle\neq0$.
However, we are principally interested in the fields ${\phi_{i}}$,
which have zero expectation value in the absence of $Z_{4}$ symmetry-breaking.
The mean-field theory for the action (\ref{K-matrix}) is the set
of coupled saddle-point equations -- the minimum with respect to fluctuating
fields $\phi_{i}$ and the maximum with respect to $\psi$.

\begin{figure}
\includegraphics[width=0.95\columnwidth]{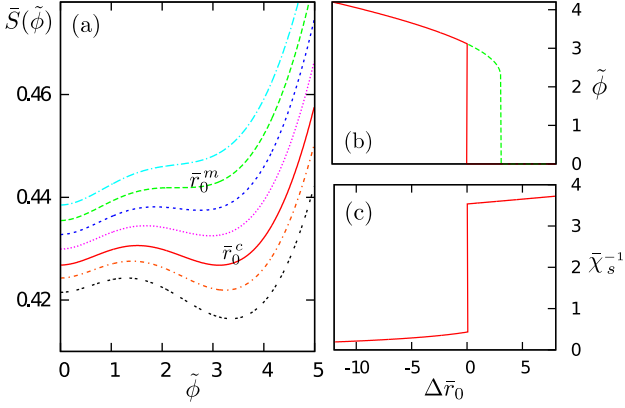} \caption{(Color online) (a) The effective action $\bar{S}(\tilde{\phi})=S[r,\tilde{\phi}]/V\bar{g}$
as a function of $\tilde{\phi}$ for $\lambda={\bar{u}}/{\bar{g}}=100$
and various $\bar{r}_{0}$. The different curves correspond to $\bar{r}_{0}=197$,
195.94 ($\bar{r}_{0}^{\, m}$), 195, 194, 192.9 ($\bar{r}_{0}^{\, c}$),
192, and 191 (from top to bottom). (b) The (red) solid curve shows
the order parameter $\tilde{\phi}$ as a function $\Delta\bar{r}_{0}=\bar{r}_{0}-\bar{r}_{0}^{\, c}$.
The (green) dashed curve shows the expectation value $\tilde{\phi}$
of the metastable phase for $\bar{r}_{0}<\bar{r}_{0}^{\, m}$. (c)~The
inverse susceptibility of the singlet mode $1/\chi_{s}\propto\bar{r}-2\tilde{\phi}$
as a function of $\Delta\bar{r}_{0}$. \label{fig:S-eff}}
\end{figure}

\subsection{Mean-field theory}

The four possible realizations for the $Z_{4}$ symmetry breaking
correspond to $\phi_{i}=\pm\phi$ subject to the constraint $\phi_{1}\phi_{2}\phi_{3}>0$.
We substitute this in Eq.~(\ref{S}), integrate over $d^{2}q$, and
absorb the factor $T$ into the couplings. We obtain
\begin{eqnarray}
 &  & \frac{S[r,\phi]}{3V}=-\frac{(r-r_{0})^{2}}{6u}+\frac{3\phi^{2}}{6g}+\frac{3r}{8\pi}\label{Seff-4}\\
 &  & \quad+\frac{1}{8\pi}\left[(r-2\phi)\log\frac{\Lambda^{2}}{r-2\phi}+2(r+\phi)\log\frac{\Lambda^{2}}{r+\phi}\right],\nonumber
\end{eqnarray}
 where $\Lambda$ is the upper momentum cutoff, and $r_{0}$, which
is proportional to the temperature, is the control parameter. The
renormalized $r$ is proportional to the inverse magnetic correlation
length 
 $\xi^{-2}$. Differentiating Eq.~(\ref{Seff-4}) with respect to
$r$ and $\phi$ yields the coupled saddle-point equations which determine
their values at a given $r_{0}$:
\begin{eqnarray}
r=r_{0}+\frac{\bar{u}}{2}\log\frac{\Lambda^{6}}{(r-2\phi)(r+\phi)^{2}},\label{eq:r}
\end{eqnarray}

\begin{eqnarray}
\phi=\bar{g}\log\left(\frac{r+\phi}{r-2\phi}\right),\label{eq:phi}
\end{eqnarray}
 where the rescaled parameters are $\bar{u}=3u/4\pi$ and $\bar{g}=g/4\pi$.
For $\phi=0$ (i.e., when $Z_{4}$ is unbroken), Eq.~(\ref{eq:r})
takes the familiar form $r+3\bar{u}/\left(2\log\Lambda^{2}/r\right)=r_{0}$
whose solution is positive for arbitrary $r_{0}$, i.e. $O(3)$ symmetry
is preserved at any non-zero $T$. This is an obvious consequence
of the Mermin-Wagner theorem. However, the discrete $Z_{4}$ symmetry
can be broken at a finite~$T$.

Assuming that $\phi$ is non-zero and eliminating $r$ from Eqs.~(\ref{eq:phi})
and~(\ref{eq:r}), we obtain the self-consistent equation for the
rescaled order parameter $\tilde{\phi}=\phi/\bar{g}$:
\begin{eqnarray}
\tilde{\phi}\left(\frac{2\mathrm{e}^{\tilde{\phi}}+1}{\mathrm{e}^{\tilde{\phi}}-1}\right)+\lambda\left[\tilde{\phi}+\frac{3}{2}\log\left(\frac{3\tilde{\phi}}{\mathrm{e}^{\tilde{\phi}}-1}\right)\right]=\bar{r}_{0},\label{r0-phi}
\end{eqnarray}
 where $\bar{r}_{0}=r_{0}/\bar{g}+(3\bar{u}/2\bar{g})\log(\Lambda^{2}/{\bar{g}})\propto(T-\bar{T}_{N})$,
and $\bar{T}_{N}$ is the rescaled mean-field $T_{N}$. The ratio
$\lambda\equiv{\bar{u}}/\bar{g}$ is large in our model, of order
$W/T_{N}$, where $W$ is the bandwidth~\cite{Nandkishore12}. The
analysis of Eq.~(\ref{r0-phi}) for $\lambda\gg1$ shows that the
first non-zero solution appears at a particular temperature when $\bar{r}_{0}^{\, m}\approx\frac{3}{2}\lambda\log3$
and at a finite $\tilde{\phi}\approx2.15+14.2/\lambda$. This obviously
indicates that the mean-field $Z_{4}$ transition is first-order.
The actual transition temperature is smaller than $\bar{r}_{0}^{\, m}$
because at $\bar{r}_{0}^{\, m}$ the effective action only develops
a local minimum at nonzero $\tilde{\phi}$, but this may not be a
global minimum. To find when the actual transition occurs, we solve
Eq.~(\ref{eq:r}) for $r(\phi)$ numerically, substitute the result
into~(\ref{Seff-4}) and obtain the effective action $S(\tilde{\phi})$
for which $\bar{r}_{0}$ is a parameter and Eq.~(\ref{eq:phi}) is
the saddle-point solution. The behavior of $S(\tilde{\phi})$ for
various $\bar{r}_{0}$ is shown in Fig.~\ref{fig:S-eff}(a). At sufficiently
large $\bar{r}_{0}$, it increases monotonically with $\tilde{\phi}$
and its only minimum is at $\tilde{\phi}=0$, implying that $Z_{4}$
is unbroken. At $\bar{r}_{0}=\bar{r}_{0}^{\, m}$, the function $S(\tilde{\phi})$
develops an inflection point, which at smaller $\bar{r}_{0}$ splits
into a maximum and a minimum. At some $\bar{r}_{0}=\bar{r}_{0}^{\, c}$
the value of $S(\tilde{\phi})$ at this minimum becomes equal to $S(0)$,
and for $\bar{r}_{0}<\bar{r}_{0}^{\, c}$, the global minimum of the
free energy jumps to a finite $\tilde{\phi}\neq0$. Once this happens,
the system spontaneously chooses one out of four states with $\pm\tilde{\phi}$,
and the $Z_{4}$ symmetry breaks down. We plot $\tilde{\phi}$ versus
$\bar{r}_{0}$ in Fig.~\ref{fig:S-eff}(b).

To find how much the $Z_{4}$ transition temperature $T_{Z_{4}}$
actually differs from the SDW transition temperature ${\bar{T}}_{N}$,
we computed the spin susceptibility $\chi(q)$ within RPA, explicitly
related ${\bar{r}}_{0}$ to $(T-{\bar{T}}_{N})$, and expressed $\lambda$
in terms of the ratio of $T_{N}$ and the fermionic bandwidth $W$.
Collecting all factors we find

\begin{equation}
T_{Z_{4}}=\bar{T}_{N}+a\frac{T_{N}^{2}}{W}\,\frac{1}{\log W/T_{N}}\label{new}
\end{equation}
where $a=O(1)$, and $T_{N}$ is the ``mean field'' Neel temperature,
which does not take into account the suppression of SDW order by thermal
fluctuations. The actual ${\bar{T}}_{N}$ tends to zero in 2D, but
$T_{Z_{4}}$ remains finite.

To analyze how the broken $Z_{4}$ symmetry affects SDW correlations,
we compute the eigenvalues of the spin susceptibility matrix $\mathcal{\hat{X}}$
in~(\ref{K-matrix}). The two eigenvalues correspond to a singlet
and a doublet mode $\chi_{s}=1/(r-2\phi)$ and $\chi_{d}=1/(r+\phi).$
If either $r-2\phi$ or $r+\phi$ jumped to a negative value at the
$Z_{4}$ transition, then the breaking of $Z_{4}$ would induce a
simultaneous breaking of the $O(3)$ symmetry. However, it follows
from (\ref{eq:phi}) that both $\chi_{s}$ and $\chi_{d}$ remain
finite when $\phi$ jumps to a nonzero value, i.e. breaking the $Z_{4}$
symmetry does not induce SDW order immediately (see Fig. \ref{fig:S-eff}(c)).

\subsection{Beyond mean-field: 4-state Potts model}

The effective action (\ref{Seff-4}) can be expanded for small $\phi$
and large $\lambda$ as:
\begin{equation}
\frac{S(\tilde{\phi})}{V}=\left(\bar{r}_{0}-\bar{r}_{0}^{\, m}\right)\tilde{\phi}^{2}-\frac{\lambda}{12}\,\tilde{\phi}^{3}+\frac{\lambda}{16}\,\tilde{\phi}^{4}+\cdots,\label{exp_S}
\end{equation}
 This action has the same form as that of the 4-state Potts model~\cite{Zia75},
implying that both transitions belong to the same universality class.
We can use this analogy to go beyond the saddle-point solution and
understand how the $Z_{4}$ transition is affected by fluctuations
of $\phi$ fields. The 4-state Potts model in 2D does exhibit a transition,
i.e. the preemptive $Z_{4}$ ordering is not destroyed by fluctuations~\cite{Wu82}.
Interestingly, however, fluctuations transform the first-order transition
into a second-order transition, although with a rather small critical
exponent $\beta=1/12$ for $\phi\sim(T_{c}-T)^{\beta}$ (Ref.~\cite{Wu82}).
A small $\beta$ implies that the order parameter sharply increases
below the critical temperature, and in practice this behavior is almost
indistinguishable from that in the first-order transition.

 Notice that the cubic term in the action, which comes from the product
$\phi_{1}\phi_{2}\phi_{3}$, ensures that the $Z_{4}$ symmetry breaking
belongs to the universality class of the 4-state Potts model, and
not of the 4-state clock model. This distinction is important, as
they have different critical behaviors in two dimensions. While the
4-state Potts model has a $\beta=1/12$ exponent, as discussed above,
the 4-state clock model transition belongs to the same universality
class of the Ising model, with $\beta=1/8$.

\subsection{Experimental manifestations}

As spin rotational symmetry is preserved in the preemptive $Z_{4}$
phase, no magnetic Bragg peaks are to be observed in neutron scattering
experiments. On the other hand, since the charge density $\rho(\mathbf{r})$
and the Casimir operator $\mathbf{S}^{2}(\mathbf{r})$ have the same
symmetry, a spatial modulation of the latter induces a modulation
in the charge density. Given the 2D character of graphene, such a
super-lattice structure can be directly probed by STM. The additional
Bragg peaks due to the quadrupled unit cell should also be detectable
by scattering measurements. Local probes such as NMR can measure the
different on-site fluctuating magnetic moments of the $Z_{4}$ phase,
since the size of the local moment controls the linewidth of the NMR
signal. We thus expect to see two different linewidths coming from
the $3\Delta$ and the $\Delta$ sites.

The order parameter $\phi$ can also be inferred by measuring the
static magnetic susceptibility $\chi$ at any of the three nesting
vectors. In the absence of $O(3)$ breaking, we have $\chi(\bar{r}_{0})=(2\chi_{d}+\chi_{s})/3$.
Once the order parameter $\phi$ jumps to a finite value below the
transition, so does the susceptibility $\chi(\bar{r}_{0})=\tilde{r}^{-1}+\phi^{2}\tilde{r}^{-3}+\cdots$,
where $\tilde{r}$ is the value of $r$ at $\phi=0$. This provides
a direct method for detecting the order parameter $\phi$. The jump
of the static susceptibility (i.e. of the spin correlation length)
also affects the electronic spectrum. For larger correlation length
the system develops precursors to the SDW order, which give rise to
a pseudogap in the electronic spectral function. This pseudogap can
be probed by photoemission experiments~\cite{fernandes}.

\section{Conclusion}

We discussed in this work the intriguing possibility of an emergent
paramagnetic phase with spontaneously broken translational symmetry
for properly doped fermions on triangular and hexagonal lattices.
This unique state emerges from a preemptive phase transition which
breaks only a discrete translational $Z_{4}$ lattice symmetry but
preserves $O(3)$ spin-rotational invariance. We demonstrated that
this phase exists in 2D systems and by continuity should exist in
anisotropic 3D systems. We argued that such a phase should be observed
in STM, NMR, neutron scattering, and photoemission experiments.

\textit{Acknowledgement.} We acknowledge useful conversations with
Leonid Levitov, Patrick Lee, Liang Fu, C.~D. Batista, Ivar Martin,
and Philip Kim. G.W.C. is supported by ICAM and NSF-DMR-0844115, R.M.F.
acknowledges the support from ICAM and NSF-DMR 0645461, as well as
the valuable support from the NSF Partnerships for International Research
and Education (PIRE) program OISE-0968226. A.V.C. is supported by
NSF-DMR-0906953.

\appendix

\section{Hubbard-Stratonovich transformation}

In this Appendix we present the details of the Hubbard-Stratonovich
transformation for the preemptive phase. We first introduce six bosonic
fields $\psi$, $\zeta_{1}$, $\zeta_{2}$, and $\phi_{i}$ ($i=1,2,3)$,
each corresponding to one of the fourth-order terms in Eq.~\ref{S_eff}
of the main text. Explicitly, the interaction terms in the partition
function $Z=\int\mathcal{D}\bm{\Delta}_{i}\exp(-S[\bm{\Delta}_{i}])$
can be rewritten as\begin{widetext}
\begin{eqnarray}
\exp\left[{-\frac{u}{2}\int_{x}\left(\bm{\Delta}_{1}^{2}+\bm{\Delta}_{2}^{2}+\bm{\Delta}_{3}^{2}\right)^{2}}\right] & = & \int\mathcal{D}\psi\,\exp\int_{x}\left[\frac{\psi^{2}}{2u}-\psi\left(\bm{\Delta}_{1}^{2}+\bm{\Delta}_{2}^{2}+\bm{\Delta}_{3}^{2}\right)\right],\\
\exp\left[{-\frac{v}{2}\int_{x}\left(\bm{\Delta}_{1}^{2}+\bm{\Delta}_{2}^{2}-2\bm{\Delta}_{3}^{2}\right)^{2}}\right] & = & \int\mathcal{D}\zeta_{1}\,\exp\int_{x}\left[\frac{\zeta_{1}^{2}}{2v}-\zeta_{1}\left(\bm{\Delta}_{1}^{2}+\bm{\Delta}_{2}^{2}-2\bm{\Delta}_{3}^{2}\right)\right],\\
\exp\left[{-\frac{v}{2}\int_{x}3\left(\bm{\Delta}_{1}^{2}-\bm{\Delta}_{2}^{2}\right)^{2}}\right] & = & \int\mathcal{D}\zeta_{2}\,\exp\int_{x}\left[\frac{\zeta_{2}^{2}}{2v}-\sqrt{3}\zeta_{2}\left(\bm{\Delta}_{1}^{2}-\bm{\Delta}_{2}^{2}\right)\right],\\
\exp\left[{\frac{g}{2}\int_{x}(\bm{\Delta}_{i}\cdot\bm{\Delta}_{j})^{2}}\right] & = & \int\mathcal{D}\phi_{k}\,\exp{\int_{x}\left[-\frac{\phi_{k}^{2}}{2g}+\phi_{k}(\bm{\Delta}_{i}\cdot\bm{\Delta}_{j})\right]},
\end{eqnarray}
\end{widetext} where $(ijk)$ in the last equation are cyclic perturmations
of $(123)$. The new action becomes
\begin{eqnarray}
S[\bm{\Delta}_{i},\psi,\bm{\zeta},\bm{\phi}] & = & \sum_{ij}\int_{q}\mathcal{X}_{ij}[\psi,\bm{\zeta},\bm{\phi}]\,(\bm{\Delta}_{i}\cdot\bm{\Delta}_{j})\nonumber \\
 & + & \int_{x}\left(\frac{|\bm{\phi}|^{2}}{2g}-\frac{|\bm{\zeta}|^{2}}{2v}-\frac{\psi^{2}}{2u}\right),\label{Seff-2}
\end{eqnarray}
 where $\bm{\zeta}=(\zeta_{1},\zeta_{2})$ and $\bm{\phi}=(\phi_{1},\phi_{2},\phi_{3})$,
and the matrix $\mathcal{\hat{X}}$ is
\begin{eqnarray}
\mathcal{\hat{X}}=\!\left(\begin{array}{ccc}
\tilde{\chi}_{q}^{-1}+2\,\hat{\mathbf{u}}_{1}\!\cdot\!\bm{\zeta} & -\phi_{3} & -\phi_{2}\\
-\phi_{3} & \tilde{\chi}_{q}^{-1}+2\,\hat{\mathbf{u}}_{2}\!\cdot\!\bm{\zeta} & -\phi_{1}\\
-\phi_{2} & -\phi_{1} & \tilde{\chi}_{q}^{-1}+2\,\hat{\mathbf{u}}_{3}\!\cdot\!\bm{\zeta}
\end{array}\right),\quad\label{K-matrix2}
\end{eqnarray}
 with $\tilde{\chi}_{q}^{-1}=r_{0}+\psi+q^{2}\equiv r+q^{2}$, and
the three unit vectors are $\hat{\mathbf{u}}_{1,2}=\left(1/2,\,\pm\sqrt{3}/2\right)$,
and $\hat{\mathbf{u}}_{3}=(-1,0)$. Integrating out the $\bm{\Delta}_{i}$
fields yields an effective action
\begin{eqnarray}
S[\psi,\bm{\zeta},\bm{\phi}] & = & \frac{3}{2}\int_{q}\log\bigl(\det\mathcal{\hat{X}}[\psi,\bm{\zeta},\bm{\phi}]\bigr)\nonumber \\
 & + & \int_{x}\left(\frac{|\bm{\phi}|^{2}}{2g}-\frac{|\bm{\zeta}|^{2}}{2v}-\frac{\psi^{2}}{2u}\right),\label{eq:S_eff3}
\end{eqnarray}
 As discussed in the main text, the mean-field solution of the potential
preemptive phase is given by the saddle-point solution of coupled
equations: $\partial S/\partial\psi=\partial S/\partial\bm{\zeta}=\partial S/\partial\bm{\phi}=0$.
In particular, we consider the two equations involving the doublet
$\bm{\zeta}$:
\begin{eqnarray}
\zeta_{1} & = & \frac{3v}{2}\int_{q}\frac{6\zeta_{2}^{2}-6\zeta_{1}\left(\tilde{\chi}_{q}^{-1}+\zeta_{1}\right)-\left(\phi_{1}^{2}+\phi_{2}^{2}-2\phi_{3}^{2}\right)}{\det\mathcal{\hat{X}}},\nonumber \\
\zeta_{2} & = & \frac{3v}{2}\int_{q}\frac{6\zeta_{2}\left(2\zeta_{1}-\tilde{\chi}_{q}^{-1}\right)-\sqrt{3}\left(\phi_{1}^{2}-\phi_{2}^{2}\right)}{\det\mathcal{\hat{X}}}.
\end{eqnarray}
 It can be easily checked that the mean-field configurations with
$\zeta_{1}=\zeta_{2}=0$ and $|\phi_{1}|=|\phi_{2}|=|\phi_{3}|=\phi$
are solutions of the above two equations, indicating that the $Z_{4}$
phase solutions discussed in the main text satisfy the saddle-point
equations of the effective action~(\ref{eq:S_eff3}).

\end{document}